# Eye dominance and testing order effects in the circularly-oriented macular pigment optical density measurements that rely on the perception of structured light-based stimuli

Mukhit Kulmaganbetov,[1,2,3,]* Taranjit Singh,[1,2] Dmitry Pushin,[1,4,5,6] Pinki Chahal,[7] David Cory,[4,8] Davis Garrad,[5] Connor Kapahi,[4,5] Melanie Mungalsingh,[9] Iman Salehi,[9] Andrew Silva,[9] Ben Thompson,[1,2,9] Zhangting Wang,[1] Dusan Sarenac[1,6,7,9]

[1] Centre for Eye and Vision Research, 17W Hong Kong Science Park, Hong Kong
[2] Entoptica Limited, 136 Bonham Strand, Sheung Wan, Hong Kong
[3] Kazakh Eye Research Institute, Almaty, Kazakhstan, A05H2A8
[4] Institute for Quantum Computing, University of Waterloo, Waterloo, ON, Canada, N2L3G1
[5] Department of Physics, University of Waterloo, Waterloo, ON, Canada, N2L3G1
[6] Incoherent Vision Inc., Wellesley, ON, Canada, N0B2T0
[7] Department of Physics, University at Buffalo, State University of New York, Buffalo, New York 14260, USA
[8] Department of Chemistry, University of Waterloo, Waterloo, ON, Canada, N2L3G1
[9] School of Optometry and Vision Science, University of Waterloo, Waterloo, ON, Canada, N2L3G1
*mukhit.k@cevr.hk

## Abstract

Psychophysical discrimination of structured light (SL) stimuli may be useful in screening for various macular disorders, including degenerative macular diseases. The circularly-oriented macular pigment optical density (coMPOD), calculated from the discrimination performance of SL-induced entoptic phenomena, may reveal a novel functional biomarker of macular health. In this study, we investigated the potential influence of eye dominance and testing order effects on SL-based stimulus perception, factors that potentially influence the sensitivity of screening tests based on SL technology.

A total of 28 participants (aged 18-38 years) were selected for the study after undergoing a comprehensive eye examination. A psychophysical task was performed where various SL-based entoptic images with multiple azimuthal fringes rotating with a specific temporal frequency were projected onto the participants' retinas. By occluding the central areas of entoptic images, we measured the retinal eccentricity $R_T$ of the perceivable area of the stimuli. The slope of the coMPOD profile ($a$-value) was calculated for each participant using a spatiotemporal sensitivity model that takes into account the perceptual threshold measurements of structured light stimuli with varying spatial densities and temporal frequencies. The mean±SD $a$-values for the dominant and non-dominant eyes were 0.11°±0.06° and 0.11°±0.05°, respectively. Similarly, the values for the first and second eyes were 0.11°±0.05° and 0.10°±0.05°, respectively. The Pearson correlation coefficient between eye dominance and testing order effects was r=0.8 (p<0.01). The Bland-Altman plots for both factors indicated zero bias.

The results indicate repeatable measurements for both eyes, implying minimal impact from eye dominance and testing order on SL-based stimulus perception. The results provide a foundation for future studies exploring the clinical utility of SL tools in eye health.

**Intro**

The perception of polarised light has attracted significant interest in vision science due to its potential applications in detecting and diagnosing macular diseases [1,2]. Standard methods, such as Haidinger's brush (HB), have been limited in effectiveness due to weak signals, low contrast, restricted stimulus flexibility, and a narrow visual extent [2–6]. Using structured light (SL) techniques, light can be engineered to have distinct spatial polarisation states that elicit rich entoptic patterns and mitigate these limitations [7–9]. Notably, SL-induced entoptic phenomena with azimuthal fringes $N_f \geq 11$ can be observed beyond the foveal region, extending up to $\approx 5°$ retinal eccentricity $R_T$ [7] compared to HB with $R_T \approx 2°$ [1,2,4,6].

SL-based stimulus perception has the potential to be used as a diagnostic tool [3,10,11] for detecting early functional signs of macular degenerative diseases such as age-related macular degeneration (AMD) [12–14], type 2 macular telangiectasia (MacTel) [15–17], and pathologic myopia [18–20].

The ability to perceive polarisation-induced entoptic phenomena is due to the presence of dichroic macular pigments (MP) [4,21], which are cumulated in the inner plexiform (IPL), outer nuclear (ONL) and Henle fibre layers (HFL) of the retina [22–24]. By modulating the spatial density ($N_f$) and temporal frequency ($\omega$) of the azimuthal fringes and occluding areas ($R_T$) in SL-based entoptic images, it is possible to characterise the circularly-oriented macular pigment optical density (coMPOD) in the radially-oriented fibres of photoreceptors in HFL [25].

SL-based tests are monocular, and clinical applications may involve comparisons between the two eyes in the same observer. Therefore, it is important to assess whether factors such as eye dominance and testing order on the test results. Eye dominance can involve oculomotor control [26,27] and cortical processing [28–31]. However, an individual's dominant eye may not always align with their optimal ability to perceive polarisation [32,33].

Studying the impact of eye testing order aids in identifying sources of variability and determining the technique's reliability and consistency [34–36]. The assessment of SL-based stimulus perception in a monocular context [3,7–11,25,37] raises the possibility that the observer may exhibit task learning from one test to the next. Thus, it is important to ascertain potential learning effects.

**Methods**

*Study Participants*

A total of 35 participants (mean age 25±5.5 years, range 18-38 years, 17F/18M) were recruited from the Centre for Eye and Vision Research (CEVR). Each participant underwent a comprehensive eye and vision examination prior to inclusion, which included assessments of medical and family history, unaided visual acuity, subjective refraction using trial frames and lenses, binocular vision tests, ocular motility, contrast sensitivity using the Thomson Chart (Thomson Software Solutions, UK), slit-lamp biomicroscopy, indirect ophthalmoscopy, colour fundus photography (Nidek AFC-330, Japan), optical coherence tomography (Topcon DRI OCT Triton, Japan), ocular biometry (Zeiss IOLMaster 700, Germany), and macular pigment optical densitometry (MPS II, UK). Eye dominance was determined with the Miles test. Only participants with good vision (best-corrected visual acuity = 0.00 logMAR or better in each eye) and healthy eyes were included (n=28), excluding those with a history of eye disease, strabismus, ocular injury, arterial hypertension, or neurological disorders (n=7).

This cross-sectional study (ClinicalTrials.gov ID: NCT05913063) adhered to the Declaration of Helsinki principles and received ethical approval from the Hong Kong Polytechnic University

(HSEARS20210910002). Detailed study objectives and procedures were communicated to all participants, who provided written informed consent prior to data collection.

*Psychophysical Task*
The top row of Figure 1 depicts various target stimuli when viewed through an ideal radial polariser. During the study, various SL-based entoptic images with visible fringe numbers $N$ rotating with a specific temporal frequency $\omega$ for each $N_f$ ($N_f$=7 with $\omega$=1.9 Hz; $N_f$=12 with $\omega$=3.8 Hz; $N_f$=17 with $\omega$=5.8 Hz; $N_f$=22 with $\omega$=7.7 Hz), were projected onto the retina of each participant for 0.5 seconds per trial. The contrast of the pattern is perceived to be highest near the centre and diminishes towards the outer edges due to the decreasing concentration of circularly-oriented macular pigments with retinal eccentricity [25]. The entoptic profile was rotated either clockwise or counterclockwise using a motorised polariser. Participants identified the rotation direction of the entoptic pattern within a two-alternative forced choice task. A spatial light modulator (SLM) created a circular obstruction at the centre of the entoptic image (Figure 1, bottom). Participants were instructed to fixate on a guide light generated by aligning the centre of the stimuli illuminated by a red laser with a maximum retinal eccentricity of 10 pixels (≈0.45°).

Similar to our previous studies [7,25], we used SL techniques to occlude central areas of different sizes in the generated entoptic images to determine their retinal eccentricity $R_T$. This threshold mask initially started from a radius in visual degrees of 10 pixels (≈0.45°), and if a participant could detect the correct direction of rotation of the stimuli, it increased with a step-size of 30 pixels (≈1.35°). A 2-up/1-down staircase method aimed to estimate the size of each entoptic image that produced 71% discrimination accuracy [37]. Participants with more than one reversal point (floor performance) at the minimum obstruction radius (≈0.45°) were excluded from the analysis.

Participants removed any refractive correction, such as spectacles or contact lenses, before the experiment, which was conducted monocularly. Both eyes were tested on the same day with a short break (1-2 mins) between eyes. The first eye to be examined was chosen randomly, with the contralateral eye covered. The second eye was tested under the same conditions.

*Circularly-Oriented Macular Pigment Optical Density*
The spatiotemporal sensitivity model introduced in ref [25] for characterising the optical density of the circularly-oriented macular pigment showed that coMPOD profile $M(r)$ is inversely proportional to the retinal eccentricity $r$ and can be described by the equation:

$$M(r) = \frac{a}{2\pi r} \qquad (1)$$

where the $a$-value is determined based on the perceptual threshold measurements of structured light stimuli with different spatiotemporal frequencies and provides a unique characterization tool for macular health. It serves as the decay constant for the radial polarizer efficiency in the human eye, directly quantifying the amount of circularly-oriented macular pigment present throughout the region of interest. This parameter, influenced by the number of Henle's fibres and the efficiency of the radial polarisation effect of the contained macular pigment, is essential for quantifying the visual extent of various spatiotemporal SL stimuli and individualising the $a$-value, as detailed in the model introduced in [25].

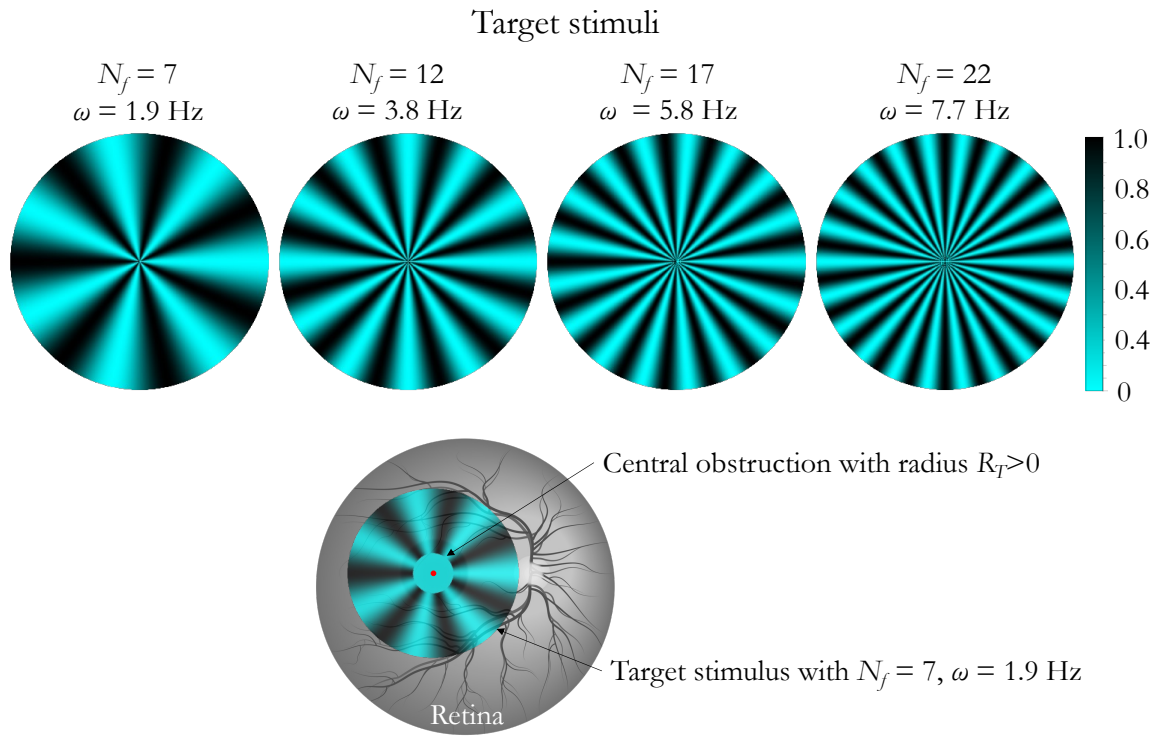

Figure 1. Target stimuli (top) with $N_f$=7, 12, 17 and 22 with $\omega$=1.9, 3.8, 5.8 and 7.7 Hz, respectively, as viewed through an ideal polariser. *Enface* projection of the entoptic pattern with a central obstruction with radius $R_T$ and $N_f$=7 azimuthal fringes onto the retina (bottom)

*Statistical Analysis*

Interocular differences in functional (visual acuity, contrast sensitivity) and anatomical (HFL volume, anterior chamber depth, axial eye length, central corneal and crystalline length thicknesses) indicators were compared using the paired t-test. $a$-value for each eye of each participant was calculated using the threshold radius of the central obstruction for SL-induced entoptic patterns with ($N_f$=7, $\omega$=1.9 Hz), ($N_f$=12, $\omega$=3.8 Hz), ($N_f$=17, $\omega$=5.8 Hz), ($N_f$=22, $\omega$=7.7 Hz), corresponding to an angular velocity $2\pi\omega/N_f$ and was used to compute the Pearson correlation coefficient between dominant (DE) and non-dominant (NDE) eyes, as well as between the first (FE) and second (SE) eyes. The Bland-Altman method was used for the visualisation of the agreement and bias between measurements from the DE vs NDE and FE vs SE by plotting the difference between the measurements against the average of the measurements.

**Results**

There were no statistically significant differences observed in various ocular health indicators among the study participants (Table 1). No differences in the best-corrected visual acuity in the spherical and cylindrical aberrations were noted. This homogeneity in the baseline measures ensures that potential confounding factors are minimised, thereby strengthening the validity and reliability of the study findings.

Table 1. Summary of interocular differences

| Indicator | Right eye | Left eye | Paired t-test | |
|---|---|---|---|---|
| | mean ± SD | | t | p-value |
| Visual acuity, logMAR | 0.24 ± 0.33 | 0.19 ± 0.30 | 1.69 | 0.103 |
| Contrast sensitivity, $\log_{10}$ | 1.68 ± 0.15 | 1.71 ± 0.14 | -1.3 | 0.201 |
| HFL volume, $mm^3$ | 2.44 ± 0.21 | 2.45 ± 0.19 | -0.41 | 0.685 |
| MPOD | 0.42 ± 0.16 | 0.41 ± 0.13 | 0.49 | 0.629 |
| AEL, mm | 24.41 ± 1.27 | 24.34 ± 1.22 | 1.43 | 0.166 |
| CCT, mm | 0.54 ± 0.04 | 0.53 ± 0.04 | 1.66 | 0.109 |
| ACD, mm | 3.39 ± 0.36 | 3.4 ± 0.35 | 0.04 | 0.968 |
| CLT, mm | 3.86 ± 0.46 | 3.84 ± 0.46 | 1.57 | 0.128 |

*Abbreviations: ACD – anterior chamber depth; AEL – axial eye length; CCT – central corneal thickness; CLT – crystalline length thickness; HFL – Henle fire layer; MPOD – macular pigment optical density; SD – standard deviation*

Of 28 participants who completed the task, twenty-two participants were right-eye dominant, and six were left-eye dominant. The average of threshold radii ($R_T$) for the SL-based entoptic phenomena with $N_f$=7, 12, 17, and 22 were presented in Table 2 for the DE and NDE, whereas Table 3 shows the data for the testing order for the FE and SE. The results indicate minimal variances in the average central obstruction size, which suggests a high degree of consistency and similarity in the measurements obtained for both eyes.

Table 2. Central obstruction radius $R_T$ for the various SL-based entoptic images with visible fringe numbers $N_f$=7, 12, 17, and 22 for the dominant and non-dominant eyes (DE & NDE)

| Number of visible fringes (n of eyes) | Dominant eye | Non-dominant eye |
|---|---|---|
| | $R_T$ mean ± SD [°] | |
| $N_f$=7 ($n_{DE}$=16, $n_{NDE}$=19) | 2.235 ± 0.085 | 3.284 ± 0.089 |
| $N_f$=12 ($n_{DE}$=28, $n_{NDE}$=27) | 4.087 ± 0.099 | 4.299 ± 0.098 |
| $N_f$=17 ($n_{DE}$=$n_{NDE}$=27) | 5.836 ± 0.089 | 6.598 ± 0.103 |
| $N_f$=22 ($n_{DE}$=$n_{NDE}$=28) | 7.546 ± 0.099 | 7.072 ± 0.102 |

Table 3. Central obstruction radius $R_T$ for the various SL-based entoptic images with visible fringe numbers $N_f$=7, 12, 17, and 22 for the first and second eyes (FE & SE)

| Number of visible fringes (n of eyes) | First eye | Second eye |
|---|---|---|
| | $R_T$ mean ± SD [°] | |
| $N_f$=7 ($n_{FE}$=17, $n_{SE}$=18) | 2.762 ± 0.088 | 2.844 ± 0.088 |
| $N_f$=12 ($n_{FE}$=27, $n_{SE}$=28) | 4.222 ± 0.101 | 4.161 ± 0.098 |
| $N_f$=17 ($n_{FE}$=$n_{SE}$=27) | 6.194 ± 0.084 | 6.239 ± 0.108 |
| $N_f$=22 ($n_{FE}$=$n_{SE}$=28) | 7.648 ± 0.101 | 6.971 ± 0.1 |

*Interocular characteristics of coMPOD model*

In line with the findings of reference [25], our study involved adjusting the spatial density and temporal frequency ($N_f$=7, $\omega$=1.9 Hz; $N_f$=12, $\omega$=3.8 Hz; $N_f$=17, $\omega$=5.8 Hz; $N_f$=22, $\omega$=7.7 Hz) with the central obstruction radius $R_T$ (from Tables 2 and 3) of the SL entoptic pattern to calculate the coMPOD profile. The mean±SD parameters for individual participant fits of the $a$-value for eye dominance revealed an average of 0.11°±0.06° for the DE and 0.11°±0.05° for the NDE.

Additionally, the data for testing order effects showed that for the FE and SE, the values were 0.11°±0.05° and 0.10°±0.05°, respectively. In both interocular variables, coMPOD $a$-values suggest no significant learning effect for the SE and no effect of eye dominance.

Considering the role of the $a$-value in the characterisation of coMPOD profile $M(r)$ in Eq. 1, we used this variable to determine the correlation of interocular difference in the perception of structured light-based stimuli both for eye dominance and testing order effects.

A total of 26 participants (52 eyes) were included in the correlation and Bland-Altman methods; two participants were excluded from the analysis due to the inability to accurately model their binocular results, as they had fewer retinal eccentricity thresholds for either of their eyes. Pearson correlation coefficient (r) for both interocular factors (eye dominance and testing order effects) was 0.8 with p<0.01 (Figure 2). The Bland-Altman plot (Figure 3) depicts the agreement between the measurements of $a$-value in the DE vs NDE and FE vs SE. The red dashed line represents the mean difference between the eyes, which is close to 0, indicating a minimal bias in the measurements. The upper and lower limits of agreement (green dashed lines) are approximately equal to one SD (0.06). The majority of data points fall within the limits of agreement, suggesting good concordance of the interocular $a$-value.

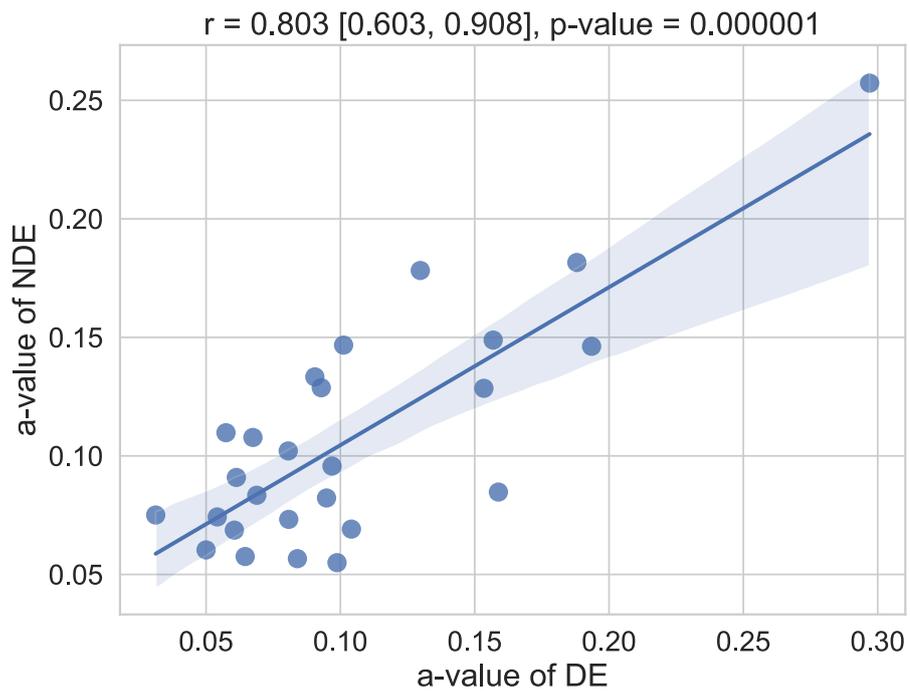

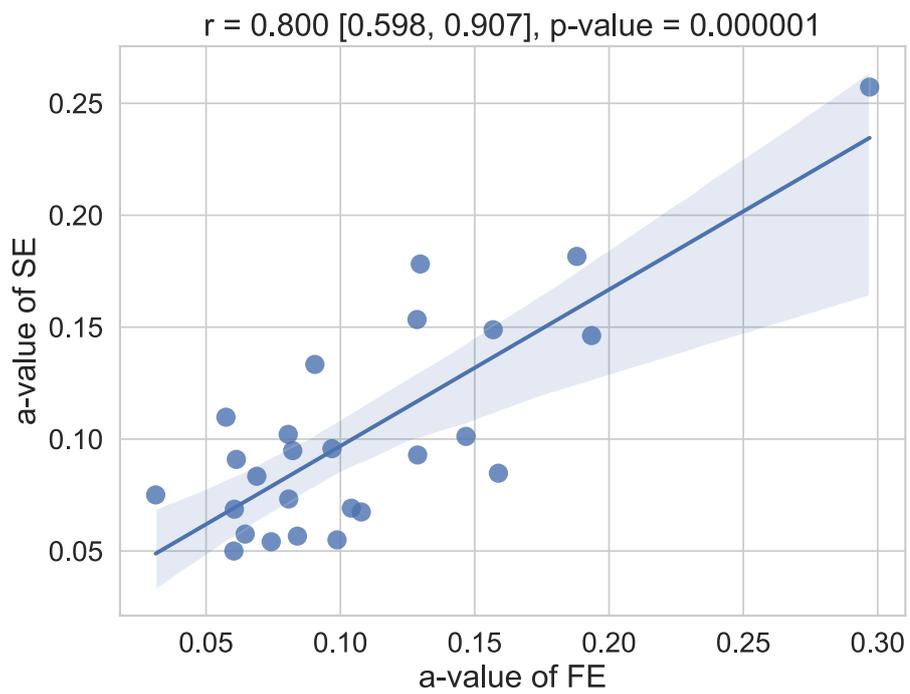

Figure 2. Correlation coefficient (r), confidence interval (CI) of the beta value of regression line and p-value of the $a$-values of dominant (DE) vs non-dominant (NDE) eyes (top) and first (FE) vs second (SE) eyes (bottom)

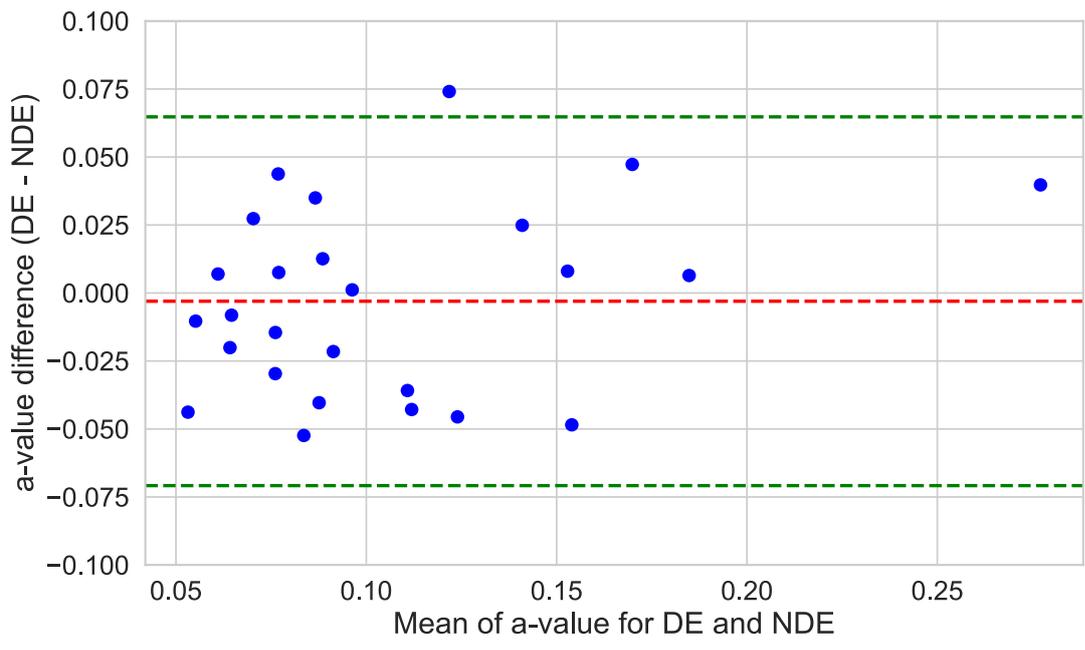

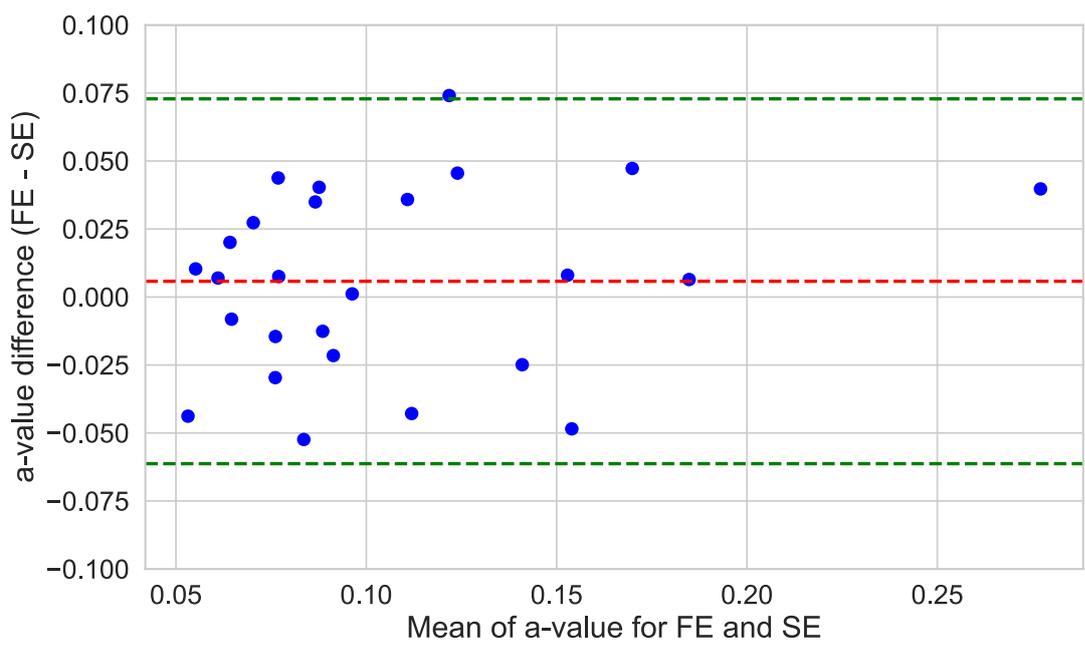

Figure 3. Bland-Altman plot for dominant (DE) vs non-dominant (NDE) eyes (top) and first (FE) vs second (SE) eyes (bottom): interocular difference measurements against their mean, with the red dashed line representing the mean difference. The green dashed lines indicate the upper and lower limits of agreement, calculated as 1.96 times the standard deviation of the differences

**Discussion**

The perception of structured light stimuli in both of an individual's eyes with similar interocular health indicators (visual acuity, contrast sensitivity, macular volume of Henle fibre layer, macular pigment optical density and ocular biometry parameters) is invariant to ocular dominance [32,33], and eye testing order. In addition, the eye dominance and testing order correlation plots displayed a strong positive correlation between the interocular $a$-value. Based on these results, it is appropriate to compare the measured coMPOD from each eye of a patient with a unilateral or asymmetric retinal condition [38–42]. These parameters of SL-based stimulus perception present a promising prospect for the future utilisation of this phenomenon in the assessment of the coMPOD profile in patients with various macular pathologies.

Although sufficient for initial observations, the sample size may not capture the full range of interindividual variability in SL-based stimulus perception. Future studies with larger and more clinically diverse populations are necessary to validate these findings. In addition, investigating polarisation perception in patients with macular disorders could provide further insights into the diagnostic potential of SL-based techniques. Another area for future research involves exploring the underlying mechanisms driving the observed consistency in polarisation perception. Understanding how the retinal polariser influences SL-based stimulus perception could enhance the development of more precise and targeted diagnostic tools.

The promising prospects for further deploying the entoptic phenomenon as an early detection tool of macular diseases such as AMD, MacTel type 2 and pathologic myopia stem from its ability to provide objective, quantitative and selective measurements of macular pigments. By utilising this phenomenon, clinicians and researchers can potentially identify subtle changes in the macula associated with macular degeneration at a much earlier stage than traditional diagnostic methods allow. This could lead to improved patient outcomes by enabling timely interventions and treatments [43].

**Conclusion**

In conclusion, the study demonstrates that eye dominance and eye testing order do not significantly influence the perception of SL-induced entoptic images. The robustness of structured light-based stimulus perception highlights the promising prospects for further deployment of SL in early detection tools for macular pathologies.

# References

1. Misson GP, Anderson SJ, Armstrong RA, Gillett M, Reynolds D. The Clinical Application of Polarization Pattern Perception. Transl Vis Sci Technol. 2020 Oct 28;9(11):31. doi: 10.1167/tvst.9.11.31.
2. Temple, S. E., Roberts, N. W., & Misson, G. P. (2019). Haidinger's brushes elicited at varying degrees of polarization rapidly and easily assesses total macular pigmentation. Journal of the Optical Society of America A: Optics, Image Science, and Vision, 36(4), B123–B131
3. Pushin DA, Cory DG, Kapahi C, Kulmaganbetov M, Mungalsingh M, Silva AE, Singh T, Thompson B, Sarenac D. Structured light enhanced entoptic stimuli for vision science applications. Front Neurosci. 2023 Jul 25;17:1232532. doi: 10.3389/fnins.2023.1232532.
4. Mottes J, Ortolan D, Ruffato G. Haidinger's brushes: Psychophysical analysis of an entoptic phenomenon. Vision Research. Volume 199. 2022. 108076. https://doi.org/10.1016/j.visres.2022.108076
5. Philipp L Müller, Simone Müller, Martin Gliem, Kristina Küpper, FrankGHolz, Wolf M Harmening, and Peter Charbel Issa, "Perception of Haidinger brushes in macular disease depends on macular pigment density and visual acuity," Investigative Ophthalmology & Visual Science 57, 1448–1456 (2016).
6. Temple SE, McGregor JE, Miles C, Graham L, Miller J, Buck J, Scott-Samuel NE, Roberts NW. Perceiving polarization with the naked eye: characterization of human polarization sensitivity. Proc Biol Sci. 2015 Jul 22;282(1811):20150338. doi: 10.1098/rspb.2015.0338.
7. Kapahi C, Silva AE, Cory DG, Kulmaganbetov M, Mungalsingh MA, Pushin DA, Singh T, Thompson B, Sarenac D. Measuring the visual angle of polarization-related entoptic phenomena using structured light. Biomed Opt Express. 2024 Jan 30;15(2):1278-1287. doi: 10.1364/BOE.507519.
8. Pushin D., Kapahi C., Silva A., Cory D., Kulmaganbetov M., Mungalsingh M., et al.. (2024). Psychophysical discrimination of radially varying polarization entoptic phenomena. Phys. Rev. Applied 21, L011002
9. Sarenac D, Andrew E Silva, Connor Kapahi, DG Cory, B Thompson, and Dmitry A Pushin, "Human psychophysical discrimination of spatially dependant pancharatnam–berry phases in optical spin-orbit states," Scientific Reports 12, 3245 (2022)
10. Sarenac D, Connor Kapahi, Andrew E Silva, David G Cory, Ivar Taminiau, Benjamin Thompson, and Dmitry A Pushin, "Direct discrimination of structured light by humans," Proceedings of the National Academy of Sciences 117, 14682–14687 (2020).
11. Sarenac D, Kapahi C, Cory DG, Pushin DA. Preparing a structured optical beam for human observation. US Patent 11,564,562, 2023
12. Ayoub T, Patel N. Age-related macular degeneration. J R Soc Med. 2009 Feb;102(2):56-61. doi: 10.1258/jrsm.2009.080298.
13. Chaudhuri M, Hassan Y, Bakka Vemana PPS, Bellary Pattanashetty MS, Abdin ZU, Siddiqui HF. Age-Related Macular Degeneration: An Exponentially Emerging Imminent Threat of Visual Impairment and Irreversible Blindness. Cureus. 2023 May 29;15(5):e39624. doi: 10.7759/cureus.39624.
14. Ayoub T, Patel N. Age-related macular degeneration. J R Soc Med. 2009 Feb;102(2):56-61. doi: 10.1258/jrsm.2009.080298.
15. Kedarisetti KC, Narayanan R, Stewart MW, Reddy Gurram N, Khanani AM. Macular Telangiectasia Type 2: A Comprehensive Review. Clin Ophthalmol. 2022 Oct 10;16:3297-3309. doi: 10.2147/OPTH.S373538.

16. Charbel Issa P, Gillies MC, Chew EY, Bird AC, Heeren TF, Peto T, Holz FG, Scholl HP. Macular telangiectasia type 2. Prog Retin Eye Res. 2013 May;34:49-77. doi: 10.1016/j.preteyeres.2012.11.002. Epub 2012 Dec 3. PMID: 23219692; PMCID: PMC3638089.
17. Pauleikhoff L, Heeren TFC, Gliem M, Lim E, Pauleikhoff D, Holz FG, Clemons T, Balaskas K; MACTEL STUDY GROUP; Egan CA, Charbel Issa P. Fundus Autofluorescence Imaging in Macular Telangiectasia Type 2: MacTel Study Report Number 9. Am J Ophthalmol. 2021 Aug;228:27-34. doi: 10.1016/j.ajo.2021.03.022
18. Takashi Ueta, So Makino, Yuuka Yamamoto, Harumi Fukushima, Shigeko Yashiro, and Miyuki Nagahara, "Pathologic myopia: an overview of the current under- standing and interventions," Global health & medicine 2, 151–155 (2020)
19. Ohno-Matsui K, Wu PC, Yamashiro K, Vutipongsatorn K, Fang Y, Cheung CMG, Lai TYY, Ikuno Y, Cohen SY, Gaudric A, Jonas JB. IMI Pathologic Myopia. Invest Ophthalmol Vis Sci. 2021 Apr 28;62(5):5. doi: 10.1167/iovs.62.5.5. Erratum in: Invest Ophthalmol Vis Sci. 2021 Jun 1;62(7):17. doi: 10.1167/iovs.62.7.17.
20. Luo N, Wang Y, Alimu S, Zhao L, Huang Y, Guo Z, Zhao X, Liu B, Chen S, Lu L. Assessment of Ocular Deformation in Pathologic Myopia Using 3-Dimensional Magnetic Resonance Imaging. JAMA Ophthalmol. 2023 Aug 1;141(8):768-774. doi: 10.1001/jamaophthalmol.2023.2869.
21. Misson GP, Temple SE, Anderson SJ. Polarization perception in humans: on the origin of and relationship between Maxwell's spot and Haidinger's brushes. Sci Rep. 2020 Jan 10;10(1):108. doi: 10.1038/s41598-019-56916-8.
22. Obana A, Nakazawa R, Noma S, Sasano H, Gohto Y. Macular Pigment in Eyes With Macular Hole Formation and Its Change After Surgery. Transl Vis Sci Technol. 2020 Oct 26;9(11):28. doi: 10.1167/tvst.9.11.28
23. Gupta AK, Meng R, Modi YS, Srinivasan VJ. Imaging human macular pigments with visible light optical coherence tomography and superluminescent diodes. Opt Lett. 2023 Sep 15;48(18):4737-4740. doi: 10.1364/OL.495247. PMID: 37707890
24. Bernstein PS, Delori FC, Richer S, van Kuijk FJ, Wenzel AJ. The value of measurement of macular carotenoid pigment optical densities and distributions in age-related macular degeneration and other retinal disorders. Vision Res. 2010 Mar 31;50(7):716-28. doi: 10.1016/j.visres.2009.10.014.
25. Pushin, D. A., Garrad, D. V., Kapahi, C., Silva, A. E., Chahal, P., Cory, D. G., ... & Sarenac, D. (2024). Characterizing the circularly-oriented macular pigment using spatiotemporal sensitivity to structured light entoptic phenomena. arXiv preprint arXiv:2409.04416.
26. Lopes-Ferreira D, Neves H, Queiros A, Faria-Ribeiro M, Peixoto-de-Matos SC, González-Méijome JM. Ocular dominance and visual function testing. Biomed Res Int. 2013;2013:238943. doi: 10.1155/2013/238943
27. Shneor E, Hochstein Sh. Effects of eye dominance in visual perception. International Congress Series, Volume 1282, 2005, Pages 719-723. https://doi.org/10.1016/j.ics.2005.05.006.
28. von der Heydt Rüdiger. Visual cortical processing—From image to object representation. Frontiers in Computer Science 5. 2023 doi:10.3389/fcomp.2023.1136987
29. J. J. Fahrenfort, H. S. Scholte, V. A. F. Lamme; The spatiotemporal profile of cortical processing leading up to visual perception. Journal of Vision 2008;8(1):12. https://doi.org/10.1167/8.1.12.

30. Webster MA. Visual Adaptation. Annu Rev Vis Sci. 2015 Nov 1;1:547-567. doi: 10.1146/annurev-vision-082114-035509. Epub 2015 Oct 22.
31. Horton JC, Fahle M, Mulder T, Trauzettel-Klosinski S. Adaptation, perceptual learning, and plasticity of brain functions. Graefes Arch Clin Exp Ophthalmol. 2017 Mar;255(3):435-447. doi: 10.1007/s00417-016-3580-y. Epub 2017 Jan 14.
32. Liu S, Zhao B, Shi C, Ma X, Sabel BA, Chen X, Tao L. Ocular Dominance and Functional Asymmetry in Visual Attention Networks. Invest Ophthalmol Vis Sci. 2021 Apr 1;62(4):9. doi: 10.1167/iovs.62.4.9.
33. Le Floch A, Ropars G, Enoch J, Lakshminarayanan V. 2010 The polarization sense in human vision. Vis. Res. 50, 2048 – 2054. (doi:10.1016/j.visres.2010.07. 007)
34. Lima MA, Pagliuca LMF, Nascimento JCD, Caetano JÁ. Comparing Interrater reliability between eye examination and eye self-examination 1. Rev Lat Am Enfermagem. 2017 Oct 19;25:e2966. doi: 10.1590/1518-8345.1232.2966. PMID: 29069269; PMCID: PMC5656336.
35. Ava K. Kiser, Derek Mladenovich, Fariba Eshraghi, Debra Bourdeau, Gislin Dagnelie; Reliability and Consistency of Dark-Adapted Psychophysical Measures in Advanced Eye Disease. Invest. Ophthalmol. Vis. Sci. 2006;47(1):444-452. https://doi.org/10.1167/iovs.04-1146.
36. Lukas Recker, Christian H. Poth; Test–retest reliability of eye tracking measures in a computerized Trail Making Test. Journal of Vision 2023;23(8):15. https://doi.org/10.1167/jov.23.8.15.
37. Andrew E. Silva, Connor Kapahi, David G. Cory, Mukhit Kulmaganbetov, Melanie Mungalsingh, Taranjit Singh, Benjamin Thompson, Dmitry A. Pushin, Dusan Sarenac; Psychophysical and image-based characterization of macular pigment using structured light. Journal of Vision 2023;23(9):5907. https://doi.org/10.1167/jov.23.9.5907.
38. Trivizki O, Wang L, Shi Y, Rabinovitch D, Iyer P, Gregori G, Feuer W, Rosenfeld PJ. Symmetry of Macular Fundus Features in Age-Related Macular Degeneration. Ophthalmol Retina. 2023 Aug;7(8):672-682. doi: 10.1016/j.oret.2023.03.016.
39. Gangnon RE, Lee KE, Klein BE, Iyengar SK, Sivakumaran TA, Klein R. Severity of age-related macular degeneration in 1 eye and the incidence and progression of age-related macular degeneration in the fellow eye: the Beaver Dam Eye Study. JAMA Ophthalmol. 2015 Feb;133(2):125-32. doi: 10.1001/jamaophthalmol.2014.4252
40. Weiss AH. Unilateral high myopia: optical components, associated factors, and visual outcomes. Br J Ophthalmol. 2003 Aug;87(8):1025-31. doi: 10.1136/bjo.87.8.1025
41. Venkatesh R, Nahata H, Reddy NG, Mishra P, Mangla R, Yadav NK, Chhablani J. Is Type 2 Macular Telangiectasia a Bilateral and Symmetrical Disease Entity? J Curr Ophthalmol. 2023 Apr 29;34(4):428-435. doi: 10.4103/joco.joco_68_22
42. Joachim N, Colijn JM, Kifley A, Lee KE, Buitendijk GHS, Klein BEK, Myers CE, Meuer SM, Tan AG, Holliday EG, Attia J, Liew G, Iyengar SK, de Jong PTVM, Hofman A, Vingerling JR, Mitchell P, Klaver CCW, Klein R, Wang JJ. Five-year progression of unilateral age-related macular degeneration to bilateral involvement: the Three Continent AMD Consortium report. Br J Ophthalmol. 2017 Sep;101(9):1185-1192. doi: 10.1136/bjophthalmol-2016-309729.
43. Galindo-Camacho RM, Blanco-Llamero C, da Ana R, Fuertes MA, Señoráns FJ, Silva AM, García ML, Souto EB. Therapeutic Approaches for Age-Related Macular Degeneration. Int J Mol Sci. 2022 Oct 4;23(19):11769. doi: 10.3390/ijms231911769